\def\apj{{ApJ}}
\def\apjs{{ApJS}}

\documentclass[cup5b]{caps}
\usepackage{graphicx}
\usepackage{amssymb}
\usepackage{ociwsymp4e}  

\HeadText{S. Lopez et al.}  

\DeclareMathAlphabet{\mathsc}{OT1}{cmr}{m}{sc}
\def\testbx{bx}%
\DeclareRobustCommand{\ion}[2]{%
\relax\ifmmode
\ifx\testbx\f@series
{\mathbf{#1\,\mathsc{#2}}}\else
{\mathrm{#1\,\mathsc{#2}}}\fi
\else\textup{#1\,{\mdseries\textsc{#2}}}%
\fi}

\def\kms{km~s$^{-1}$}
\def\cm2{cm$^{-2}$}
\def\hkpc{$~h_{50}^{-1}$ kpc}
\def\icm{cm$^{-2}$}
\def\civ{\ion{C}{iv}}
\def\hi{\ion{H}{i}}
\def\nhi{$N_{\rm \ion{H}{i}}$}
\def\nciv{$N_{\rm CIV}$}
\def\lya{Ly$\alpha$}
\def\lyb{Ly$\beta$}
\def\civhi{\civ/\hi}
\def\la{\mathrel{\hbox{\rlap{\hbox{\lower3pt\hbox{$\sim$}}}{\lower-2pt\hbox{$<$}}}}}
\def\ga{\mathrel{\hbox{\rlap{\hbox{\lower3pt\hbox{$\sim$}}}{\lower-2pt\hbox{$>$}}}}}

\begin{document}

\pagenumbering{arabic}


%
\author[]{S. LOPEZ$^{1}$, S. D'ODORICO$^{2}$, S. L.  ELLISON$^{2,3}$ and
  T.-S. KIM$^{4}$ 
\\
(1) Universidad de Chile, Chile\\
(2) European Southern Observatory\\
(3) Pontificia  Universidad Cat\'olica de Chile, Chile \\
(4) Institute of Astronomy, UK
}

\chapter{
Clues to the origin of metals \\ in the Ly$\alpha$ forest}

\begin{abstract}

We analyze the $z=2$ \lya\ and \civ\ forest along two adjacent
lines-of-sight at high S/N and spectral resolution to assess the
question of the origin of metals in the intergalactic medium. We find
that the gas containing detectable amounts of metals (as seen in
absorption by \ion{C}{iv}) is less quiescent on scales of a few kpc
than the gas without metals.  Although it is difficult to discriminate
unequivocally between possible sources of enrichment and the extent to
which each contributes to line-of-sight differences,
%
those suggest that the metal-enriched systems are bound to galaxies
(i.e., they do not arise in the IGM nor in galactic winds).

\end{abstract}

\section{Introduction}


The high-$z$ IGM is inhomogeneously distributed on large-scales in
  voids and filaments; it contains a large fraction of baryons (Rauch,
  Haehnelt \& Steinmetz 1997) and is highly ionized
  and warm ($\sim 10^{4}$ K). Despite a low neutral fraction, a large
  range of densities ---from slightly underdense gas up to the denser
  environment of collapsed objects--- can be probed via \hi\
  absorption of photons from background QSOs. This is called the \lya\
  forest. The optical depth of the \hi, $\tau_{\rm Ly\alpha}$, maps
  overdensities in the low density regime. The higher the \hi\ column
  density, the stronger and kinematically more complex the system.

High \nhi\ absorbers, \nhi$\ga 10^{16}$ \icm, contain heavy metals
  (the \civ\ $\lambda\lambda 1548,1550$ \AA\ doublet is the most
  common and easy-to-recognize metal absorption feature). At low-$z$,
  strong \civ\ systems arise in galactic halos (e.g., Chen et
  al. 2001; Heckman et al. 2000). Low \nhi\ absorbers, on the other
  hand, with \nhi$\la 10^{14}$ \icm, contain no metals, or we lack of
  sensitivity to detect them (Cowie \& Songaila 1998; Ellison at
  al. 2000).  A fraction of {\it intermediate} \nhi\ absorbers,
  however, with $10^{14}\la $\nhi$\la 10^{16}$ \icm, does show weak \civ\
  at a level of $W\sim$ a few m\AA\ (Cowie et al. 1995). For these
  absorbers simple photoionization models have given a metallicity of
  $Z=0.1-1$ \% solar (Carswell, these proceedings).
 
Here we asses the question of the origin of such metals. Possible and 
  non-exclusive models  depend on when the metals were released to the IGM:
   \begin{enumerate}
       \item{Supernova-driven outflows from early galaxies with a
       small contribution from PopIII stars (e.g., Scannapieco,
       Ferrara \& Madau 2002); however, only a fraction of the IGM can
       be metal-enriched through this mechanism (but see Ferrara,
       Pettini, \& Shchekinov 2000).}
       \item{ 'In situ' or more recent  enrichment by galactic
       winds (Theuns et al. 2002); however,  metals are 
       difficult to transport over large distances.} 
     \end{enumerate}

Our aim in this work is to check the different enrichment hypotheses
    by observing the \lya\ and \civ\ forests along two adjacent
    lines-of-sight (LOSs). The idea is to study the possible
    observational effects that feedback from star formation might have
    in the two spectra.  One difficulty inherent to this approach is
    that the gas in the \lya\ forest is mostly ionized, meaning that
    either the gas metallicity $Z$,  or the density $n_{\rm H}$ is a free
    parameter when modeling the 
    ionization balance. In addition, the information on \hi\ is
    inaccurate for \nhi $\ga10^{14.5}$ \icm\ because the \lya\ lines become
    saturated.


\section{Results}


\subsection{Data}

For the first time we have obtained simultaneous coverage of the \lya\
and \civ\ forests at high S/N and resolution (VLT-UVES) along two
closely separated LOSs (the gravitationally lensed QSO
HE1104$-$1805). The redshift range for \lya\ {\it and} \civ\ is $
z=1.6$ to $ z=2.3$.  The quality of the spectra allows us to detect
\ion{C}{iv} systems down to \nciv $=10^{11.8}$ \icm. The observed
distribution of systems per unit column density, when compared with
that of Ellison et al. (2000), implies that our sample is complete down
to \nciv $=10^{12.3}$ (A) and $10^{12.5}$ (B) \icm. Typical separation
between LOSs is $2$ \hkpc. At this separation, we expect totally
correlated \lya\ forest spectra (Smette et al. 1995).

  \begin{figure}
    \centering
\includegraphics[width=7cm,angle=0]{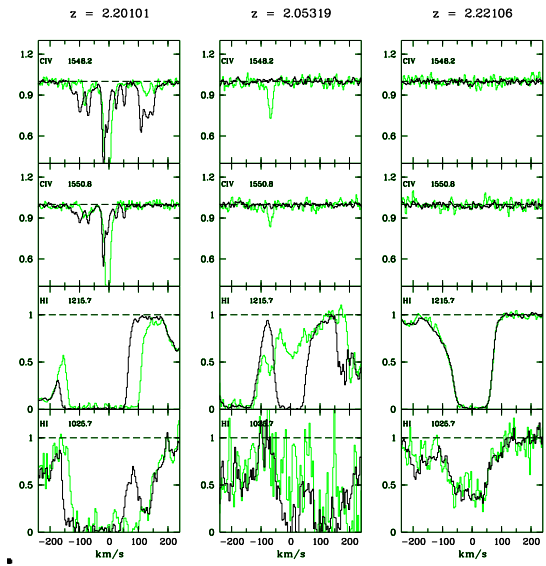}
\caption{Examples of absorption systems in
  HE1104$-$1805 A (black) and B (grey). {\it Left:}  a 'complex' metal
  system with several 
  velocity components. {\it Middle:} a 'single component' metal system observed
  in B only. {\it
  Right:} a saturated \hi\ system 
  with no \civ\  detected. Separation between LOS is only $0.6$
  \hkpc. \hi\ column densities (left to right): \nhi\
  $>9.3\times 10^{16}$, $1.0\times 10^{15}$, and $3.0\times 10^{14}$
  \cm2}    
\label{fig_CIV} 
  \end{figure}

\subsection{Approach 1: line profile fitting}

All lines in the \lya\ forest were profile-fitted with VPFIT
  automatically, and self-consistent solutions for 872 \lya\ systems
  ---half of them also with simultaneous fits to \lyb--- were found.
  At each $z_{\rm HI}$ we searched for \civ, finding 93 \civ\ doublets
  in either spectrum.  The \hi\ and \civ\ fit processes were kept
  totally independent, which is important with regard to line
  positions (a redshift offset between \civ\ and \hi\ is not
  unexpected; Ellison et al. 2000). Since many velocity components can
  be present (Fig.~\ref{fig_CIV}), we attempted an automatic
  association of \civ\ with a certain \hi\ system by summing up all
  VPFIT column densities within a given velocity window. This
  procedure also minimizes the effects of line blending and redshift
  offsets.  If \civ\ was present within $\pm \Delta$ \kms\ of $z_{\rm
  HI}$, we made the  distinction between (1) complex,
  strong and clustered systems
  and (2) single, weak and isolated systems.

\begin{figure}
    \centering
\includegraphics[width=6.5cm,angle=0]{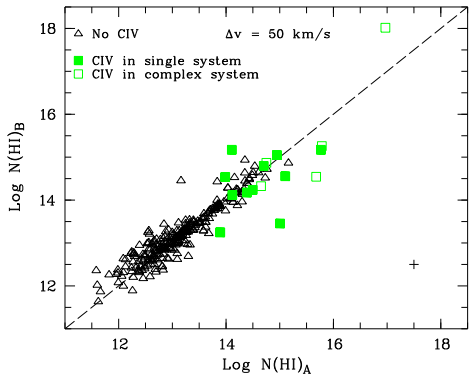}
\caption{\hi\ column density of \lya\ forest systems in HE1104$-$1805  
  A and B, with \civ\ in either or both spectra (gray squares) and
  without \civ\ 
  (black triangles). Open squares are for velocity components within
  complex systems; 
  filled squares are for 'single' systems. The cross indicates
  the typical $1\sigma$ error.}  
\label{fig_HAHB} 
  \end{figure}

Fig.~\ref{fig_HAHB} shows a comparison of \nhi\ along both LOSs for
  $\Delta = 50$ \kms. We observe that \nhi\ shows in general stronger
  variations across the LOSs in \hi+\civ\ systems than in \hi-only
  systems within the same range of \nhi\ (the scatters of the column
  density differences are respectively $\sigma_{\rm HI+CIV}=0.7$ dex
  and $\sigma_{\rm HI}=0.4$ dex ). Note that any effect of disturbed
  gas will be less evident here because possible velocity shifts are
  not considered when comparing column densities.  Therefore, {\it the
  general trend seems to be that of \nhi\ varying more strongly across
  the LOSs when \civ\ is present.}

\begin{figure}
    \centering
\includegraphics[width=12cm,angle=0]{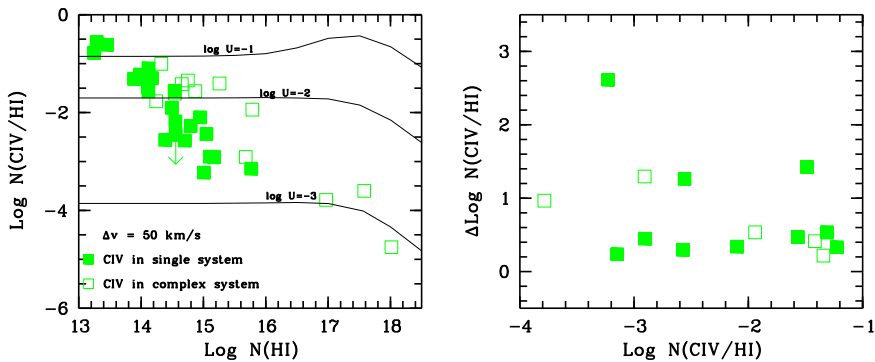}
\caption{{\it Left:} \civ/\hi\ ratio vs. \nhi\ for all systems (A and B) along
  with CLOUDY photoionization models for $Z = 0.1$ solar, a
  Haardt \& Madau (1996) ionizing background spectrum and different
  ionization parameters $U$. {\it Right:} variation of the \civ/\hi\ ratio
  across the LOSs as a function of \civ/\hi.
}  
\label{fig_CLOUDY} 
  \end{figure}

If we consider every \civ\ system, we find (Fig.~\ref{fig_CLOUDY},
left panel) that single and complex systems both span a similar range
of \civ/\hi\ ratios and that, moreover, there is little distinction
between both classes. The \civ/\hi\ ratio is a function of both gas
density (thus of ionization) and metallicity, but the correlation
between \civhi\ and \hi\ that is observed here rather points out to
density variations (ionization) as the main source of
spread. Ionization parameters in the range $\log U=-3$ to $\log U=-1$
reproduce the observed range of \civ/\hi\ values for $Z=0.1$
solar. Altogether, {\it complex and single systems seem to share
similar physical properties and thus to have a common origin}.

If we investigate the {\it transverse} differences in \civhi\ between
LOSs we find that they do not correlate with \civhi\ (right panel of
Fig.~\ref{fig_CLOUDY}). Under the assumption of homogeneous
metallicity, the lack of a trend may be connected 
to the mechanism of metal transport. Consider, for example,  the observed
metals as having been produced in far galaxies and transported by
strong galactic outflows that {\it ionize} their surroundings (Adelberger
et al. 2003; Kollmeier et al. 2002). Then, the stronger the wind, the
more ionized the gas (higher \civhi), and ---simultaneously--- the more
perturbed the medium (larger $\Delta$\civhi). Therefore, and despite
of the small sample, it seems that the observed \civ\ is {\it
not} part of strong galactic outflows.

\subsection{Approach 2: Pixel counting}

\begin{figure}
    \centering
\includegraphics[width=6.5cm,angle=0]{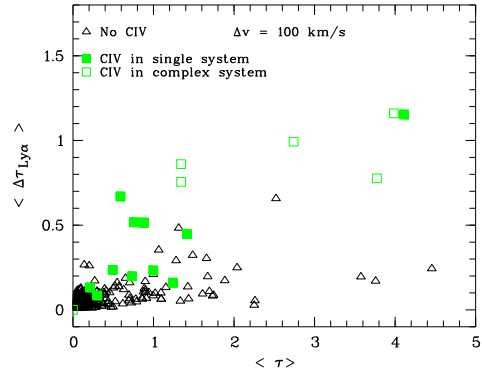}
\caption{Optical depth differences $\Delta\tau_{\rm Ly\alpha}$
  between spectra within $\pm 100$ \kms\ of $z_{\rm Ly\alpha}$ as a
  function of $\tau$. }
\label{fig_pix1} 
  \end{figure}

As an alternative to column densities, the optical depth per pixel
    $\tau_{\rm Ly\alpha}=-ln(F)$, where $F$ is the normalized flux,
    was calculated for each spectrum.  Differences $\Delta\tau$
    between the A and B spectra, and their median values
    $<\Delta\tau>$ were calculated over $\pm \Delta$ \kms\ around each
    $z_{\rm Ly\alpha}$.  Fig.~\ref{fig_pix1} shows $<\Delta \tau>$ as
    a function of $\tau$ for $\Delta = 100$ \kms\ (to account for
    slightly saturated lines, the velocity window was larger than in
    1.2.2). Mock spectra constructed to evaluate the accuracy of the
    $\Delta \tau$ measurement indicate that $\Delta \tau$ is
    confidentially well constrained if $\tau\la 3$. Also here we observe that
    $\tau_{\rm Ly\alpha}$ has more structure around \hi+\civ\ systems
    than around \hi-only systems. We argue that in this plot the
    effect is more striking than in Fig.~\ref{fig_HAHB} because of
    radial velocity shifts in systems that contain \civ.

\section{Conclusions}

Our conclusions are twofold. First, the \hi\ at $z=2$ seems to be more
  inhomogeneous if it hosts \civ, on scales for which the forest is
  otherwise featureless. Velocity shear appears to be main driver for
  the observed differences but changes in ionization might be also
  significant. Secondly, we find that weak \civ\ systems in the
  intermediate \nhi\ forest do not differ much from single components
  in more complex \civ\ systems.

As a consequence, the weak \civ\ observed here can not simply be the
  relic of an early widespread episode of metal enrichment. Rather,
  the cross-talk between both \hi\ spectra on such small scales is
  naturally explained by metal-enriched gas that is perturbed because
  it is bound to galaxies. While this is not a surprise for \nhi\
  $=10^{16}$ \icm, it is interesting to see that the trend continues
  down to \nhi\ $=10^{14}$ \icm\ absorbers. Our finding builds on
  observations of Lyman-break galaxies capable of enriching and
  ionizing the IGM (Pettini et al.2001; Adelberger et al. 2002), and
  of \civ\ absorbers having been disturbed recently (Rauch, Sargent \&
  Barlow 2001). However, it does not fit into hydrodynamical simulations that
  predict \civ-transporting galactic winds to have little effect on
  \hi\ (Theuns et al. 2002).

The similar \civ/\hi\ ratios in weak and strong systems is
  intriguing. Indeed, we may be simply lacking sensitivity to detect
  more structure along the line of sight. Thus, the possibility
  appears that the \civ\ in {\it moderate} column-density \lya\
  systems at $z=2$ has a real 'in-situ' origin, occurring in the
  extended halos of an as yet undetected population of
  galaxies. Certainly this does not exclude the possibility 
  that some of the detected systems might arise in the
  'true' IGM. An increase of sensitivity and hence in the statistics
  of weak systems will help unveiling this longstanding problem.



\end{document}